\documentstyle[times,pramana,epsf,floats]{ias}
\begin{document}
\begin{flushright} 
TIFR/TH/07-03\\
hep-ph/0702109 \\ 
\end{flushright} 
\vskip 65pt 
\begin{center} 
{\Large \bf Physics Beyond the Standard Model and Cosmological
Connections: A Summary from LCWS 06 }\\
\vspace{8mm} 
{\large\bf 
         K.~Sridhar\footnote{sridhar@theory.tifr.res.in}
}\\ 
\vspace{10pt} 
{\sf Department of Theoretical Physics, \\
                     Tata Institute of Fundamental Research,\\  
                     Homi Bhabha Road, 
                     Bombay 400 005, India. } 

\end{center} 
 
\vfill 
\clearpage

\title{Physics Beyond the Standard Model and Cosmological
Connections: A Summary from LCWS 06}

\author{K. Sridhar}
\address{Department of Theoretical Physics, 
Tata Institute of Fundamental Research,
Homi Bhabha Road, Mumbai 400 005, India}
\pacs{2.0}
\abstract{The International Linear Collider (ILC) is likely to provide
us important insights into the sector of physics that may supersede
our current paradigm viz., the Standard Model. In anticipation of 
the possibility that the ILC may come up in the middle of the next
decade, several groups are vigourously investigating its potential
to explore this new sector of physics. The Linear Collider Workshop
in Bangalore (LCWS06) had several presentations of such studies
which looked at supersymmetry, extra dimensions and other exotic
possibilities which the ILC may help us discover or understand.
Some papers also looked at the understanding of cosmology that
may emerge from studies at the ILC. This paper summarises these 
presentations.}

\maketitle
\section{Introduction}
For several years, there has been a wide-spread realisation in the High Energy
Physics community of the importance of a high-energy linear collider to study 
TeV-scale physics in the scattering of electrons, positrons and photons. 
This realisation has found concrete articulation in a project now known as 
the International Linear Collider (ILC) \cite{Aguilar-Saavedra:2001rg,Abe:2001nn1,Abe:2001nn2,Abe:2001nn3,Abe:2001nn4,Abe:2001gc}. 
Due to the nature of the leptonic 
and photonic initial states, such a collider will provide
a very clean environment setting the stage for the probing the TeV scale
with remarkably high precision. Much in the same way as the LEP experiments
in CERN complemented the UA1 and UA2 experiments in understanding the
gauge sector of the Standard Model (SM), the ILC is expected to complement
the Large Hadron Collider (LHC) in the quest to understand the
nature of electroweak symmetry breaking and its ramifications 
\cite{Weiglein:2004hn}.

Moreover, compelling theoretical arguments tell us that electroweak 
symmetry-breaking also holds the key to the discovery of new physics: 
the much sought-after `physics beyond the Standard Model'. 
From a purely phenomenological
point of view, almost any kind of new physics can be expected at the high
energies that these colliders will probe. But high-energy theorists have
focussed their research on the sub-class of models which are consistent
with some of the following principles: symmetry, renormalizability,
unitarity, naturalness etc. To these principles are added consistency with 
existing experimental data and also accessibility of the theoretical models
in upcoming experiments. These guiding principles narrow down the field
considerably and the viable models that result fall into a few distinct
classes. 

Supersymmetry, or more precisely, the supersymmetric extension of the 
Standard Model is the most popular of these classes of models \cite{Nilles:1983ge,Haber:1984rc}. Indeed,
the supersymmetric extension of the Standard Model is in no way unique 
but yields a
whole class of models. Further simplifying assumptions, some based
on aesthetic considerations and some decidedly ugly but simplifying
nonetheless, are then added to arrive at a few models which have
been the subjects of vigourous investigations. One of the major
physics goals of the ILC will be a systematic investigation of
various regions of the supersymmetric parameter space in a quest
to understand the mechanism responsible for supersymmetry breaking \cite{Abe:2001nn2}.

Another class of models that has been studied as possible extensions
of the SM are models based on the idea of Technicolour. While the minimal
versions of these models are constrained by precision electroweak data,
the non-minimal versions do manage to survive all existing constraints
\cite{Chivukula:2000mb}.
Indeed, in recent years, there has been a resurgence of interest in these
models with the realisation that they can be thought of as being dual to
some TeV scale extra-dimensional models \cite{Sundrum:2005jf}. 
Models of TeV-scale extra dimensions
have opened up new avenues for theoretical speculation in high-energy
physics \cite{Arkani-Hamed:1998rs,Randall:1999ee}. 
These models provide new solutions to the hierarchy problem and
predict a whole host of model-dependent experimental signatures at the
TeV scale \cite{Giudice:1998ck,Han:1998sg}.
While the ILC may not be a discovery machine for this class
of models, it will definitely help in pinning down the specifics of
a model by providing a window within which to investigate the details
of these models \cite{Sridhar:1999gt}.

A major part of the preparation for the ILC involve theoretical studies
and simulations
of the physics that can be potentially studied at this collider
\cite{Accomando:1997wt,Dittmaier:2003sc}. Work
has been carried out for several years now by individual researchers
as well as dedicated working groups and this work includes several
investigations of supersymmetry, extra dimensions, technicolor and
other extensions of the SM. More recently, there has been considerable
progress made in understanding the interplay between collider physics
and cosmology and how collider searches for dark matter candidates in
supersymmetry and other models can lead us to a determination of
dark matter parameters and how this precision information may influence
cosmology. This paper presents a summary of the
work on Beyond Standard Model Physics and Cosmological Connections presented 
at the Linear Collider Workshop 2006 (LCWS06) \footnote{The summary of
Higgs physics, top physics, QCD studies and loop calculations can
be found in Ref. \cite{heinemeyer}.}.

\section{Supersymmetry and the ILC}
\noindent One of the major advantages of the ILC is that it will 
allow several precision
studies of TeV-scale supersymmetry, help determine masses, branching ratios
and supersymmetric parameters to a degree of accuracy that may help
pinpoint the underlying mechanism of supersymmetry breaking.

A fundamental relation in SUSY is that between the gauge and the Yukawa
coupling:
\begin{equation}
{\rm Gauge\ coupling}\ g = {\rm Yukawa\ coupling}\ \hat g
\end{equation}

Testing these relations via precise cross-section measurements is very
important and a detailed study of how to measure these couplings
at a Linear Collider was presented \cite{freitas}. For example, 
selectron production at ILC and their decays into 
neutralinos can be studied and polarisation allows us to disentangle SU(2) 
and U(1) couplings and provides a measurement of these couplings to better 
than 1\%. Things are more complicated with SU(3) at ILC because to produce 
coloured states one needs reactions like $e^+e^- \rightarrow \tilde q \tilde 
q g$. These rates turn out to be tiny unless the ILC operates at its highest
planned centre-of-mass energy of 2 TeV.

The alternative is to do a combined LHC/ILC analysis and study production
of squarks and their decays through charginos at the LHC and input values
of squark and chargino/neutralino BRs measured precisely at the ILC.
This method gives a test of the strong coupling identity down to a 4\% level.

Precision electroweak data do not yield as sensitive an
estimate of supersymmetric particle masses as they did for the top
quark or the Higgs boson because decoupling guarantees that low-energy
observables are insenstive to the effects of supersymmetric particles.
On the other hand, loop processes involving supersymmetric particles 
may still be useful if the processes are either rare or forbidden in
the SM. In the workshop, a detailed analysis of supersymmetric models was 
presented \cite{heinemeyer2}. This analysis has been carried 
out using the precision measurements of electroweak observables like
$m_t$, $M_W$ and ${\rm sin}^2 \theta_{\rm eff}$ from LEP and Tevatron, the
bound on the lightest Higgs boson mass, loop-induced
quantities like $(g-2)_{\mu}$ and ${\rm Br}(b \rightarrow s \gamma$) and
the constraints from WMAP on the cold dark matter
density $\Omega_{\chi} h^2$ and other cosmological observations.

The different SUSY models that have been studied include:
\begin{enumerate}
\item
Constrained Minimal Supersymmetric Standard Model (CMSSM): In this model, the
universality of scalar masses, gaugino masses and trilinear parameters is
assumed at the input GUT scale.

\item
Non Universal Higgs Model (NUHM): In this model, the soft SUSY-breaking scale 
masses for the two Higgs doublets are not universal. As compared to the
CMSSM, these are two additional parameters and, at low energies, may be
taken to be the Higgs mixing parameter $\mu$ and the CP-odd Higgs boson
mass, $m_A$.

\item
Very Constrained Minimal Supersymmetric Standard Model (VCMSSM): This is
a version of the CMSSM with additional relations between the soft tri- and 
bi-linear SUSY parameters: $A_0 = B_0 + m_0$.

\item
Gravitino Dark-Matter Model (GDM): This is a model with a gravitino LSP so that
the gravitino is the dark matter candidate. But this is done with an MSUGRA
framework where the gravitino mass is taken to be equal to $m_0$ at the
input GUT scale and again the soft tri- and bi-linear parameters are
related by: $A_0 = B_0 + m_0$.
\end{enumerate}

For the CMSSM, VCMSSM and GDM, it is found that $m_{1/2}$ is 
constrained to be of the order of a few 100 GeV and this suggests light 
sparticles observable at the LHC and ILC. On the other hand, $m_0$ values 
in NUHM could be considerably larger and make production
of sparticles at ILC, at least at a centre-of-mass energy of 500 GeV, difficult.

One of the most promising channels for SUSY discovery at the ILC
is $\tilde\tau$ pair production followed by the decay of the $\tilde\tau$ to a
$\tau$ and the LSP. A work reported at the meeting \cite{guchait} showed how
the measurement of the polarisation of the $\tau$ in its one-prong hadronic
decay channel will provide information
on the underlying SUSY breaking mechanism as well as a determination of
SUSY parameters. The process studied was: $e^+e^- \rightarrow \tilde 
\tau_1 \tilde \tau_1$ with $\tau_1 \rightarrow \tau \chi_1^0$.
The normalised centre-of-mass angular distribution of the $\tau$ 
to a pion (or a vector meson) and a neutrino allows a determination of 
its polarisation $P_{\tau}$.
The composition of $\chi_1^0$ determines the polarisation of the $\tau$
and it is this composition that yields model-discrimination.
In MSUGRA: $P_{\tau}=+1$,
in non-universal SUGRA: $P_{\tau}= {\rm cos}^2 \theta_{\tau} - {\rm sin}^2 
\theta_{\tau}$, in models of anomaly mediation one gets $P_{\tau}=-1$,
 while in gauge mediation $P_{\tau}= {\rm sin}^2 \theta_{\tau} - {\rm cos}^2 
\theta_{\tau}$.
Thus it is possible to distinguish different SUSY models by measuring 
$P_{\tau}$. The work also demonstrates how $P_{\tau}$ may be used in
a global fit for a model-independent determination of SUSY parameters.

Another study presented \cite{pandita} 
dealt with the implications for gaugino masses of 
renouncing the universality of gaugino masses at the GUT scale. In the
framework of $SU(5)$ GUT, for example, such non-universality may result
from the vacuum expectation value of the $F$-term of a chiral superfield
which appears in the gauge-kinetic function.

In spite of being unequal, constraints on the neutralino and chargino masses
result from the grand-unified framework.
Upper bounds on neutralino masses and sum rules for neutralino-chargino mass
relations are obtained. Not unexpectedly, these masses have significant 
dependence on the representation of the GUT group.
These relations can be probed in the decays of neutralinos and higgs at the 
ILC.

In another presentation \cite{sopczak}, the process $e^+e^- \rightarrow \tilde t_1 \tilde 
t_1 \rightarrow c \chi_1^0 c \chi_1^0$ was studied in the case where the 
stop-neutralino mass difference is small. The scenario is motivated by 
SUSY Dark Matter 
studies where the dark matter precision determination is dependent on 
the scalar mass determination through the co-annihilation process.
An Iterative Discriminant analysis method is used to weight each event in 
such a way as to optimize signal to background.

At a centre-of-mass energy of 260 GeV, for a 122 GeV stop and 107 GeV 
neutralino, 50 fb$^{-1}$ luminosity
and 50\% efficiency 560 signal and 200 background events are estimated
using the Iterative Discriminant analysis. This goes towards a precise 
determination of the stop mass and precise prediction for dark matter.

A proposal to look for dark matter at colliders was presented \cite{sefkow} in
the context of a model where the gravitino is the lightest-supersymmetric
particle (LSP) and the dark matter candidate
and the next-to-lightest supersymmetric particle 
(NLSP) is a $\tilde\tau$ with lifetimes ranging from a few seconds to several
years. This is studied by looking at the decay of the $\tilde\tau$
via $\tilde \tau_1 \rightarrow \tau \tilde G$.  At the ILC,
copious $\tilde \tau_1$ production yields a very precise determination
of the $\tilde \tau$  parameters with results such as

\begin{eqnarray}
\delta m_{\tilde\tau} &\sim& 10^{-3} \nonumber \\
\delta m_{\tilde G} &\sim& 10^{-1}
\end{eqnarray}

The analysis can be extended to case where the NLSP is any of the other
sleptons.

\section{Extra Dimensions}

One work presented \cite{okada} studied a low-scale 
supersymmetry-breaking scenario. 
SUSY-breaking in this model is mediated by low-scale gravity in a warped 
extra dimension. Due to the
warping SUSY breaking scale $\Lambda$ is brought down to the 1 -- 10 TeV
range.
Hidden sector becomes visible due to strong interaction with the visible
sector fields and
production of hidden sector fields, $X$, at collider energies becomes 
possible.
The production and decay of $X$ at LHC and ILC has been studied and the
phenomenology is very similar to radion phenomenology.
A 1 TeV ILC should probe via $e^+e^- \rightarrow ZX$ masses of X $\sim \ 1.85$
TeV.

In another presentation \cite{bhattacharyya}, 
a model of universal extra dimension, where all the
particles are free to propagate in a fifth dimension which is compactified on 
a $S_1/Z_2$ orbifold, was
considered. Kaluza-Klein (KK) number conservation in the model yields a 
lightest KK particle (LKP) which is $\gamma_1$. 
The KK excitation of the electron $E_1$ has a mass equal
to that of the $\gamma_1$ but the degeneracy is lifted by radiative corrections.
In this work, the process $e^+ e^- \rightarrow E_1^+ E_1^-$ at the ILC
energies is studied
with $E_1 \rightarrow e \gamma_1$. This decay proceeds with a 100\% BR.
It is found that KK electrons give forward-peaked events due to $t$-channel 
dominance and this implies large forward-backward asymmetries which will
be observed at the ILC. It is shown that it is possible to use
angular distributions to obtain precise information on these
KK states.

A paper which studied the modification of the standard Einstein-Hilbert action
in models of TeV-scale gravity through the addition of higher curvature terms
was also presented \cite{rizzo}. These are higher-dimensional terms which involve the
curvature tensor and their introduction results in new ghost and scalar fields
in D-dimensions. The ghost is removed from the spectrum by fine-tuning
the model parameters but KK excitations of the scalar starting around
a TeV in mass result and the phenomenology of these scalars is studied.
It turns out that the KK excitations of the scalar has weak couplings
in both the ADD and the RS models and does not alter the phenomenology
in any serious fashion. On the other hand, the higher-curvature terms
also modify the relationship between the fundamental scale in D dimensions
and the Planck scale in 4 dimensions. This leads to a modification of the
KK sectors of the ADD and the RS models and can lead to $O(1)$ modifications
of the cross-sections for graviton emission or exchange in both models.

\section{Interface with Cosmology}

Another presentation \cite{peskin} dealt with 
the issue of how to use the ILC to identify 
the dark matter particle and to infer dark matter properties
from the measured particle spectrum and cross-sections in a
Minimal Supersymmetric Standard Model study with 24 independent parameters.
The scans of the 24-dimensional space was done using the Markov Chain
Maotne Carlo Technique for 4 SUSY points (LCC1 -- LCC4) of
the LC/Cosmology study group. It was found that
masses and splittings in typical examples have errors smaller by a factor
of 3 -- 10 compared  to LHC errors.
Also polarisation at ILC may help resolve multiple solution ambiguities
prevalent at the LHC (bino-, wino-, higgsino-like LSP).

Rate of production of gamma rays in galactic halos is proportional to
dark matter annihilation. Determination of the annihilation cross-section
allows a model independent determination of dark matter density.
The study concluded that a case for a 1 TeV ILC is very strong.

In yet another study \cite{matsumoto}, it was shown that dark matter annihilation into 
$e^+ e^-$ yields cosmic positron signals 
which can be probed in experiments like Pamela and AMS-2.
If dark matter is detected in these experiments it will allow for mass 
determination in these experiments.
The signal guarantees ILC will see $\gamma+$ DM pair production leading to
spin and properties determination of dark matter at ILC.

\section{Conclusions}
In this paper, we have summarised the presentations made at the LCWS06
in Bangalore in the sessions on Beyond Standard Model Physics and
Cosmological Connections. While significant progress has been made,
there is still a lot of concentrated effort needed in pinpointing
the role of the ILC in specific model contexts. No doubt much of this
will be helped by the physics that the LHC will provide us with and
the synergy and the complementarity of these two colliders has been
the focus of several recent studies. Another theme of recent interest
has been the interplay of collider signatures of new physics with
cosmology. Clearly, there is a lot of interesting physics to look 
forward to in the coming decade.


\end{document}